\documentclass[aps,prl,final,twocolumn,letterpaper, superscriptaddress]{revtex4-2}

\usepackage{graphicx}   
\usepackage{import}                         
\usepackage{epstopdf}
\usepackage{amsmath} 
\usepackage{float} 
\usepackage{bm}
\usepackage{amssymb}
\usepackage{quotes}
\usepackage{indentfirst}
\usepackage{color}
\usepackage[utf8]{inputenc}
\usepackage{transparent}
\usepackage{dcolumn}
\usepackage{braket}
\usepackage{multirow}
\usepackage{cancel} 
\usepackage{mdframed}
\usepackage{color}
\usepackage{bm}
\usepackage{dsfont}

\usepackage{slashed}
\usepackage{graphicx}
\usepackage{amsmath}
\usepackage{siunitx}
\usepackage{hyperref}
\usepackage{cleveref}
\crefname{equation}{Eq.}{Eqs.}
\crefname{figure}{Fig.}{Figs.}
\usepackage{soul, color} 
\soulregister\ref{7}  
\soulregister\cite{7} 
\renewcommand{\st}[1]{}

\usepackage{xr}
\usepackage{subfiles}

\makeatletter
\newcommand*{\addFileDependency}[1]{
  \typeout{(#1)}
  \@addtofilelist{#1}
  \IfFileExists{#1}{}{\typeout{No file #1.}}
}
\makeatother

\usepackage{textcomp} 
\usepackage{xifthen}
\usepackage{xcolor}
\usepackage{etoolbox}
\newboolean{togglechanges} 

\setboolean{togglechanges}{false}

\newcommand{\comment}[1]{\ifbool{togglechanges}
    {#1}  
    {\textcolor{blue}{#1}}}

\usepackage{bibentry}

\usepackage{graphicx}
\usepackage{dcolumn}
\usepackage{bm}

\usepackage{textcomp} 

\makeatletter
\renewcommand{\fnum@figure}{\textbf{Fig.~\thefigure}}
\makeatother

\begin{document}
\rmfamily

\title{Separating partially coherent light}


\author{Paul-Alexis~Mor}
\affiliation{Edward L. Ginzton Laboratory, Stanford University, Stanford, CA 94305, USA}
\affiliation{Corps des Mines, Mines Paris, Universit\'e PSL, 75272 Paris, France}

\author{Anne~R.~Kroo}
\affiliation{Edward L. Ginzton Laboratory, Stanford University, Stanford, CA 94305, USA}

\author{Carson~G.~Valdez}
\affiliation{Edward L. Ginzton Laboratory, Stanford University, Stanford, CA 94305, USA}

\author{Marko~\v{S}imi\'{c}}
\affiliation{Edward L. Ginzton Laboratory, Stanford University, Stanford, CA 94305, USA}
\affiliation{University of Graz, and Christian Doppler Laboratory for Structured Matter Based Sensing, Graz 8010, Austria}

\author{Aviv~Karnieli}
\affiliation{Edward L. Ginzton Laboratory, Stanford University, Stanford, CA 94305, USA}

\author{Gabriele~Cavicchioli}
\affiliation{Edward L. Ginzton Laboratory, Stanford University, Stanford, CA 94305, USA}
\affiliation{Politecnico di Milano, Milano 20133, Italy}

\author{Zhanghao~Sun}
\affiliation{Edward L. Ginzton Laboratory, Stanford University, Stanford, CA 94305, USA}

\author{Vittorio~Grimaldi}
\affiliation{Edward L. Ginzton Laboratory, Stanford University, Stanford, CA 94305, USA}
\affiliation{Politecnico di Milano, Milano 20133, Italy}

\author{Shanhui~Fan}
\affiliation{Edward L. Ginzton Laboratory, Stanford University, Stanford, CA 94305, USA}

\author{Olav~Solgaard}
\affiliation{Edward L. Ginzton Laboratory, Stanford University, Stanford, CA 94305, USA}

\author{David~A.\,B.~Miller}
\affiliation{Edward L. Ginzton Laboratory, Stanford University, Stanford, CA 94305, USA}

\author{Charles~Roques-Carmes}
\email{crc@ista.ac.at}
\affiliation{Edward L. Ginzton Laboratory, Stanford University, Stanford, CA 94305, USA}
\affiliation{Institute of Science and Technology Austria (ISTA), Klosterneuburg 3400, Austria}



\clearpage 




\begin{abstract}
Recent advances in optical imaging and communication increasingly involve high-dimensional, partially coherent light, creating a growing need for scalable tools to measure and manipulate coherence. Here, we demonstrate the automatic separation of spatially partially coherent light into ``coherence modes''~\cite{wolf_new_1982, roques-carmes_measuring_2024} -- its orthogonal and mutually incoherent components. To make this separation possible, we exploit variational processing in layered self-configuring interferometer architectures in a silicon photonic circuit. This process formally finds and measures the eigenvectors and eigenvalues of the coherency matrix, hence measuring the partially coherent state, while leaving it intact and separated after optimization. Furthermore, we show that mutually incoherent beams, if spatially orthogonal, can be automatically separated even if they are completely overlapped, hence separating unknown laser beams based only on their mutual incoherence. Our experiment finds and separates the two strongest coherence modes starting from a nine-mode sampling of the partially or fully overlapping fields from two independent lasers. 
The method requires a number of physical components that scales linearly with the rank $r$ of the coherency matrix and operates through a sequence of $r$ \textit{in situ} gradient-based optimizations enabled by electronic drive frequency multiplexing of interferometer phase shifters. We benchmark its performance against a mixture-based tomographic method, also implemented on chip. These results establish a scalable framework for programmable coherence analysis and control in imaging, communication, and photonic information processing.
\end{abstract}
\maketitle



Since the invention of the laser, modern optics and photonics have been built around coherent light, which underpins much of optical science and technology, from interferometry to laser-based communication and computation~\cite{saleh2019fundamentals}. As a result, the emergence of partial coherence in realistic optical systems is often regarded as a detrimental effect to be mitigated or suppressed, as it degrades interference, stability, and information transfer~\cite{bourassin2015partially, zhu2002free, salem2004polarization, ghalaii2022quantum, korotkova2021non}. 
Yet a growing body of work shows that partial coherence, when engineered, can instead be harnessed as a resource, enabling enhanced performance or entirely new functionalities in imaging and sensing~\cite{cai2005ghost, redding2012speckle, lombardo2025leveraging, tsang2016quantum}, optical computing~\cite{dong_partial_2024, jia_partially_2024}, multiplexing strategies for communication~\cite{alkhazragi2023chaotic, harling2025optical,nardi2022encoding,harling2024isoentropic}, beam propagation~\cite{gbur2002spreading, dogariu2003propagation}, as well as wavefront synthesis in holography for augmented and mixed reality technologies~\cite{choi_flexible_2022, choi_synthetic_2025, peng_speckle-free_2021, lee_light_2020}. Whether treated as a nuisance or an opportunity, these applications share a common requirement: the ability to quantitatively measure, separate, process, and ultimately engineer partially coherent light fields~\cite{korotkova2020applications}. 

Nevertheless, scalable and integrated hardware capable of processing partial coherence remains absent. In statistical optics, the modal diagonalization of a coherency matrix into mutually incoherent and orthogonal eigenfunctions -- commonly referred to as ``coherence modes''~\cite{wolf_new_1982, roques-carmes_measuring_2024} -- is a standard analytical tool~\cite{goodman_statistical_2015, gamo1964iii}. However, its direct physical implementation -- whereby these coherence modes are spatially demultiplexed into distinct output ports -- has thus far remained a theoretical proposal~\cite{roques-carmes_measuring_2024}. 
Bridging this gap would allow coherence to be manipulated in hardware in the same way as polarization, wavelength, or spatial modes, which are routinely accessed and manipulated in photonic signal processing architectures~\cite{bogaerts2020programmable}.

Measuring and processing coherence entails accessing the statistical correlations between the many degrees of freedom of an optical field, encoded in its coherency matrix $\rho_{ij}=\langle E_i E_j^*\rangle$, where $(i,j)$ label electric-field components projected onto spatial, spectral, and/or polarization bases, and $\langle \cdot\rangle$ denotes statistical (e.g., time) averaging. In practice, $\rho$ is typically reconstructed using projection-based tomography, in which the field is sequentially measured in multiple bases and the resulting data are post-processed to estimate the coherency matrix numerically ~\cite{goodman_statistical_2015, kagalwala2015optical, ploschner2022spatial}. More recently, innovative programmable photonic architectures have been demonstrated to improve stability and integration of coherence measurements and to generate programmable multimode partially coherent light (or their equivalent modal Stokes parameter descriptions)~\cite{hashemi2026programmable, hashemi2026chip}. While general, such schemes can require $\mathcal{O}(N^2)$ destructive measurements (e.g, by complete absorption in photodetectors) for an $N$-dimensional field; the $\mathcal{O}(N^2)$ scaling may render such approaches impractical for the high-dimensional and dynamically evolving coherence states encountered in imaging, communication, and photonic computing systems. This inefficiency is especially pronounced given the low rank of many partially coherent fields, which, in principle, allows their coherence structure to be captured with far fewer degrees of freedom~\cite{choi_synthetic_2025, korotkova2020applications}. Furthermore, such schemes do not physically separate partially coherent light into its constituent components, nor do they leave those components available for subsequent optical processing.

\begin{figure*}
\centering
\vspace{-0.3cm}
  \includegraphics[scale=0.6]{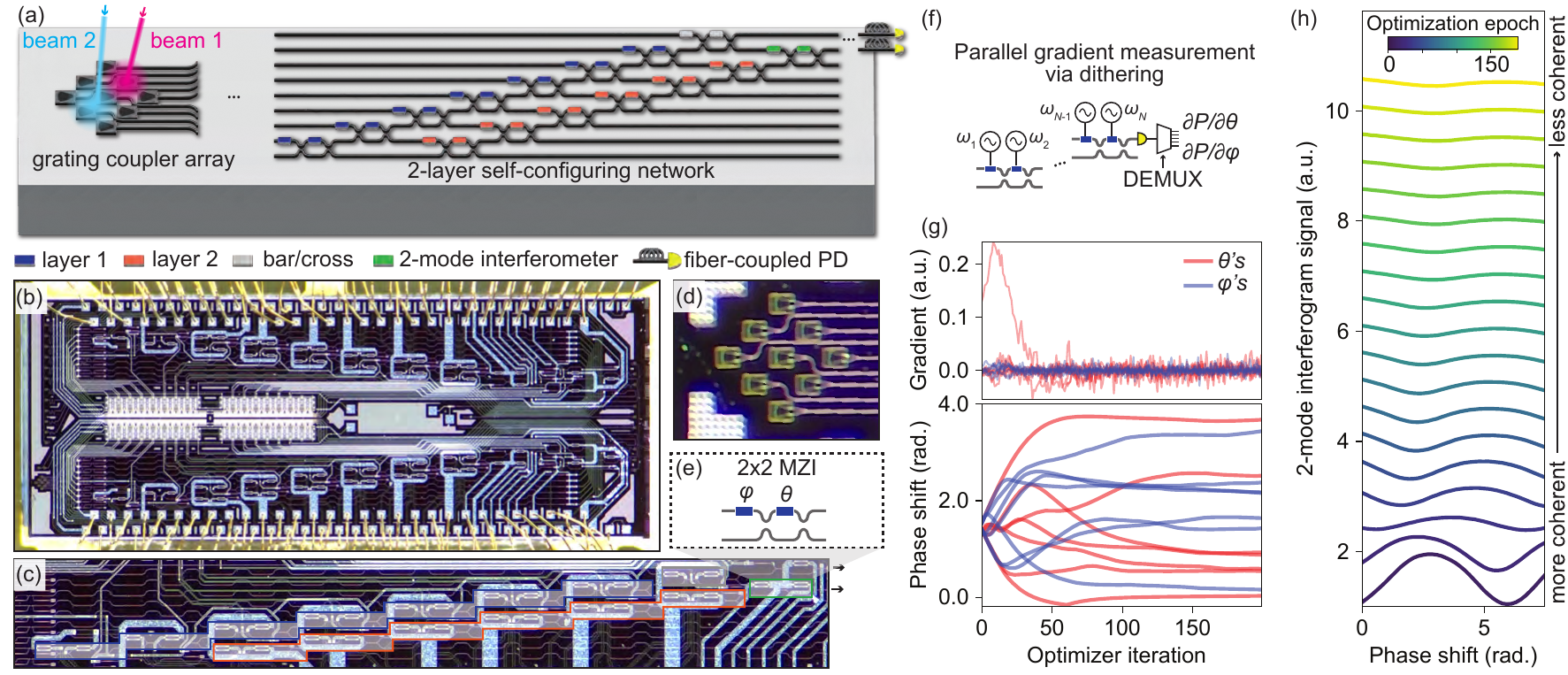}
  \caption{\small \textbf{An integrated photonic circuit to process spatial partial coherence.} (a) Schematic of the chip and incident beam layout. The chip consists of a 2-layer self-configuring network. (b) Optical micrograph of the whole chip. (c) Zoom-in on the two self-configuring layers. (d) Zoom-in on the grating coupler array, which acts as a free-space interface. (e) Schematic of a $2\times2$ Mach-Zehnder interferometer (MZI) circuit. (f) Schematic of the method for parallel gradient measurement via voltage dithering. (g) Evolution over an optimization run of the network parameters ($\theta$ and $\varphi$, bottom), and the sensitivity of the power output (first layer) to these parameters ($\partial P/ \partial \theta$, $\partial P/ \partial \varphi$). (h) Two-mode interferogram measured between the outputs of layers 1 and 2 (measured with MZI highlighted in green in (a,b). Interferograms corresponding to different epoch numbers are shifted along the $y$-axis for readability.}
    \label{fig:1}
    \vspace{-0.3cm}
\end{figure*}


Here, we experimentally demonstrate a foundry-fabricated photonic integrated platform capable of measuring, separating, processing, generating, and recording spatially partially coherent light fields, providing direct physical access to their coherency matrices. For a rank-$r$ coherency matrix in an $N$-mode space, our approach implements on-chip coherence tomography using a self-configuring interferometric network that physically separates coherence modes into distinct waveguide outputs. The tomography task is formulated as a sequence of $r$ variational optimizations (gradient descents), each extracting one coherence mode and requiring at most $\mathcal{O}(N)$ physical components. These optimizations are enabled by on-chip gradient measurements through electronic drive frequency multiplexing. Consequently, the total number of physical components scales as $\mathcal{O}(rN)$. This combined photonic hardware and gradient-based physical optimization framework constitutes a demonstration of a ``partially coherent light analyzer''~\cite{roques-carmes_measuring_2024}.

We demonstrate this approach in the spatial domain using a nine-mode device and benchmark its performance against a mixture-based tomography implemented on the same chip, achieving $\gtrsim98\%$ fidelity. We further demonstrate two proof-of-concept applications: direct measurement of the informational content of partially coherent optical fields (based on the entropy of the coherency matrix), which also enables discrimination of spatially overlapping beams; and a self-configuring network that separates orthogonal, mutually incoherent beams purely through their coherence properties. Together, these results establish an integrated and scalable platform for partially coherent light processing, paving the way toward programmable coherence-enabled photonic systems.

\section{Results}

\subsection*{Variational optimization of light with partially coherent light analyzer (PCLA)}

Our experimental setup (\cref{fig:1}(a)) interfaces free-space optical excitation with an integrated network of Mach-Zehnder interferometers (MZI) connected to a set of grating couplers. To generate arbitrary input coherency matrices with rank $r=2$, we leverage mutual (temporal) incoherence between two otherwise spatially coherent beams. Specifically, we couple in two mutually incoherent (i.e., statistically independent) optical fields at $\lambda \approx 1535$~nm. The resulting input coherency matrix is
\begin{equation}
     \rho_\text{in} = I_1| e_1\rangle\langle  e_1|+I_2| e_2\rangle\langle  e_2|,
     \label{eq:mixturedec}
\end{equation}
where \(I_i\) is the total intensity from laser \(i\in\{1,2\}\) coupled into the chip, and \(| e_i\rangle\) the normalized field amplitude distribution of that laser over the \(N\) input modes. In particular, the two modes $| e_1\rangle$ and $| e_2\rangle$ are not necessarily spatially orthogonal. 

The free-space beams are oriented approximately at the designed optimum input angle of the grating couplers to ensure efficient coupling. Light is then processed by the photonic integrated circuit (\cref{fig:1}(b)), composed of a two-dimensional array of $N=9$ grating couplers (\cref{fig:1}(d)) coupling into two self-configuring layers of MZIs (blue and orange boxes in \cref{fig:1}(c)). The two topmost waveguide outputs are connected to two fibers whose powers are read by photodetectors. Both lasers are amplitude modulated (at the same modulation frequency) and photodetector signals are read out with a lock-in amplifier. Each MZI consists of two phase shifters denoted as $(\varphi, \theta)$, thermally actuated and controlled independently by an external electronic circuit, and two 50:50 directional couplers (\cref{fig:1}(e)). The full network imparts a unitary transformation on the input light field $U_\text{PCLA}$ and transforms the coherency matrix according to $\rho_\text{out} = U_\text{PCLA} \rho_\text{in} U_\text{PCLA}^\dagger$. More details on our mathematical conventions and the experimental setup can be found in the Supplementary Information (SI), Section~S1-3.

The central idea of this work is to perform on-chip tomography through variational optimization of the output powers~\cite{roques-carmes_measuring_2024}, applied sequentially layer by layer, which physically realizes the eigendecomposition of the input coherency matrix,
\begin{equation}
\rho_\text{in} = \pi_1 \ket{u_1}\bra{u_1} + \pi_2 \ket{u_2}\bra{u_2}.
\label{eq:eigdec}
\end{equation}
Although this expression formally resembles the decomposition in \cref{eq:mixturedec}, two essential distinctions must be emphasized. The states $\ket{u_{1,2}}$ are eigenvectors of $\rho_\text{in}$ and are therefore orthogonal, $\braket{u_1|u_2}=0$, since $\rho_\text{in}$ is Hermitian. They represent \textit{spatially coherent modes} carrying optical powers $\pi_{1,2}$, given by the corresponding eigenvalues. Crucially, these coherence modes are by construction \textit{mutually incoherent}, as expressed by $\braket{u_1|\rho_\text{in}|u_2}=0$, which follows directly from the eigen-decomposition in \cref{eq:eigdec}.

This representation is exact: in our configuration, the coherency matrix has rank $r \leq 2$, as it is formed from two mutually incoherent laser beams. Importantly, although the eigenmodes $\ket{u_{1,2}}$ are orthogonal and mutually incoherent, they are generally linear combinations of both input laser states. Only when the original inputs $\ket{e_{1,2}}$ are themselves orthogonal does the diagonalization recover the individual laser beams.

\begin{figure*}
\centering
  \includegraphics[scale=0.59]{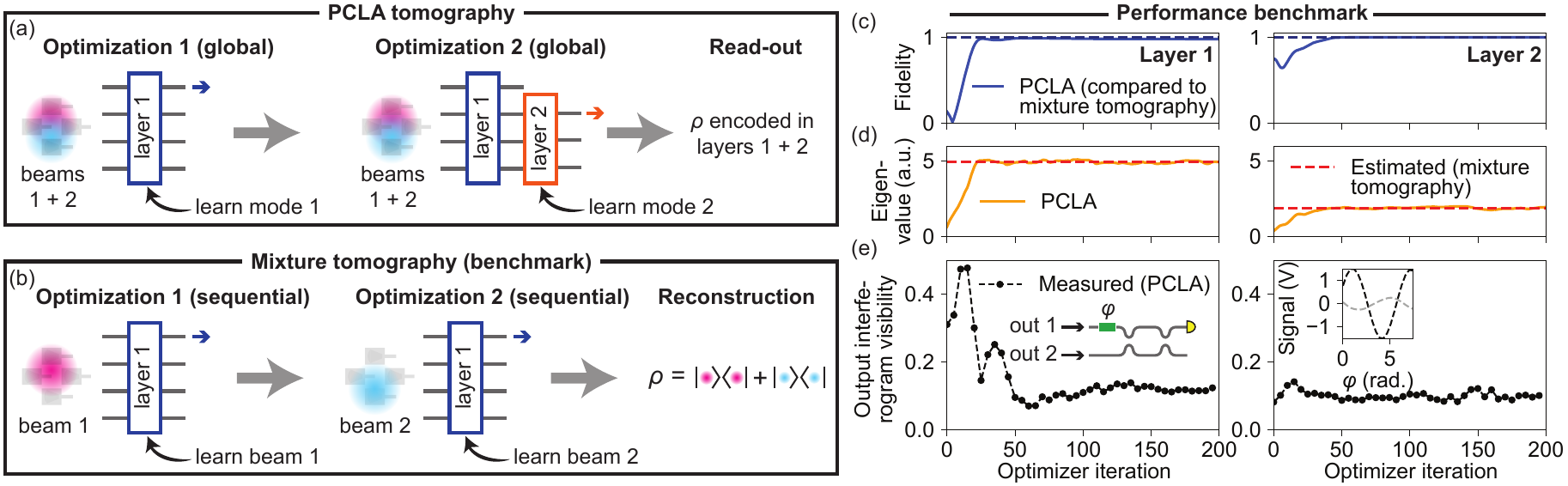}
  \caption{\small \textbf{Variational coherence tomography on an integrated photonic circuit.} (a) PCLA tomography: each coherence mode of the coherency matrix is learned by global optimization of the corresponding PCLA layer. (b) Mixture tomography: each mixture component (a spatially coherent laser beam) is learned separately, and the full coherency matrix is reconstructed numerically. (c-e) Benchmarking PCLA and mixture tomography by performing variational coherence tomography on layer 1 (left) and 2 (right). (c) Fidelity of PCLA unitary compared to ground truth unitary matrix estimated from mixture tomography. (d) Lock-in signal (corresponding to coherency matrix eigenvalue). (e) 2-mode contrast evolution over epoch number. Left inset shows a schematic of the interferometer to measure beating between outputs of layers 1 and 2. Right inset shows the interferogram for the maximum (beginning of layer 1 optimization) and minimum (layer 2 optimization) contrast positions.}
    \label{fig:2}
    \vspace{-0.3cm}
\end{figure*}

Our variational coherence tomography (\cref{eq:eigdec}) first globally optimizes all phase shifters in the first self-configuring MZI layer to maximize its output power, and then repeats this procedure for each subsequent layer (of which we implement two).
These global optimizations within a layer therefore require on-chip access to the gradients of that output power with respect to the MZI phase shifter settings $(\varphi,\theta)$ in that layer. We achieve this using a frequency-multiplexed gradient estimation scheme (\cref{fig:1}(f)), in which each phase shifter is sinusoidally dithered at a distinct frequency $\omega_i$, all chosen within a single octave. 
The photodetector time trace is demodulated against orthogonal drive signals applied to each phase shifter, yielding coefficients directly proportional to the corresponding gradients $\partial P/\partial\varphi_i$ or $\partial P/\partial\theta_i$. 
This approach enables simultaneous estimation of all gradients within a layer -- up to 14 phase shifters in the present device -- in a single measurement step. 

The extracted gradients are then supplied to a gradient-descent-based optimizer algorithm~\cite{kingma2014adam} to update the MZI parameters and maximize the output power. \cref{fig:1}(g) shows the measured gradients and the corresponding parameter evolution for a single-layer optimization, demonstrating rapid convergence to zero-mean gradients within approximately 50 iterations, consistent with convergence to the optimum. Because the optimization of output power can be formulated as a Rayleigh-quotient maximization, the convexity of the cost function guarantees that the algorithm converges to the global optimum~\cite{roques-carmes_measuring_2024, roques2025automated}. Generally, the number of iterations required to reach a given accuracy depends on the coherency matrix eigenspectrum and the measurement signal-to-noise ratio of the system, as analyzed in detail in Section~S2 of the SI.

Our chip architecture further enables interferometric access to the two output modes of the network, implemented using the MZI highlighted in green in \cref{fig:1}(c). This configuration is obtained by “dropping” the upper waveguide, achieved by setting the corresponding grey MZI to the bar state, where power from the bottom (respectively, top) input port is routed to the bottom (respectively, top) output port. The resulting interferogram exhibits fringe visibility proportional to the off-diagonal elements of the output coherency matrix $\rho_\text{out}$. As the optimization proceeds, the PCLA effectively learns the Hermitian adjoint of the eigenvector matrix $U$ with $U = (\ket{u_1}, \ket{u_2}, \star)$ (where $\star$ is a unitary matrix on the remaining modes), yielding $U_{\mathrm{PCLA}} = U^\dagger$. Consequently, the fringe visibility progressively vanishes, directly signaling the transition of the output fields from partially coherent to mutually incoherent states (\cref{fig:1}(h)). 

Importantly, the two coherence modes $\ket{u_{1,2}}$ are not merely reconstructed but are physically routed into separate output waveguides. 
Note this separation occurs entirely within a unitary interferometric transformation \textit{before arriving at the photodetectors}; so, once the device is configured, the photodetectors can be removed or switched out of the path, and the separation into coherence modes remains in these outputs. If, for example, we reflected these powers back into the outputs, the phase conjugate of the original partially coherent field would emerge from the grating couplers~\cite{MillerAnalyze2020}, and passing these outputs instead to a phase-conjugate version of the network (so, with the $\varphi$ phase shifter settings reversed) would directly create the original partially coherent field emerging from its grating couplers~\cite{miller2013reconfigurable} (see data for partially coherent light generation in the SI, Section~S5).

\subsection*{Coherence tomography in a photonic integrated circuit}

The PCLA performs coherence tomography through layer-by-layer variational optimization of the output powers~\cite{roques-carmes_measuring_2024}. Upon convergence, the measured output powers directly yield the eigenvalues $\pi_{1,2}$, while the corresponding coherence modes $\ket{u_{1,2}}$ are encoded in the optimized layer parameters $(\varphi,\theta)$, which can be extracted from the settings of the (calibrated \cite{MillerAnalyze2020}) phase shifters. To benchmark the accuracy and robustness of our approach, we implement a complementary tomography scheme based on controlled state preparation and the established ability of MZI networks to self-configure to coherent inputs \cite{miller_self-configuring_2013, MillerAnalyze2020}, which we refer to as ``mixture tomography''. In this method, individual input lasers (corresponding to mixture components) are turned on one at a time, producing reduced coherency matrices of the form $\rho_\text{in}^{(i)} = I_i \ket{e_i}\bra{e_i}$. Each coherent input state $\ket{e_i}$ is then recovered via self-configuration of the network using standard techniques~\cite{miller_self-configuring_2013, MillerAnalyze2020, milanizadeh2021coherent}, and the full coherency matrix is subsequently reconstructed numerically as $\rho_\text{in} = \rho_\text{in}^{(1)} + \rho_\text{in}^{(2)}$ (see Section~S2 of the SI). Comparing these two methods (\cref{fig:2}(c-e) implemented on our chip (here, constrained to 4 input ports), we achieve eigenvector fidelities of $97.8\%$ (layer 1) and $99.9\%$ (layer 2) and eigenvalue estimation errors of $0.29\%$ (layer 1) and $3.11\%$ (layer 2). 

Accordingly, the two-mode fringe visibility decreases during optimization of the first layer, reflecting the progressive suppression of coherence between the output modes. As expected, the fringe visibility does not decrease further during optimization of the second layer. By the end of the first layer, the network has already learned the coherence mode $\ket{u_1}$, and, by construction, the mode implemented by the second layer is constrained to be spatially orthogonal to $\ket{u_1}$, even though the corresponding optical power may not yet be fully concentrated in the second output.

\begin{figure}
\centering
\vspace{-0.3cm}
  \includegraphics[scale=0.65]{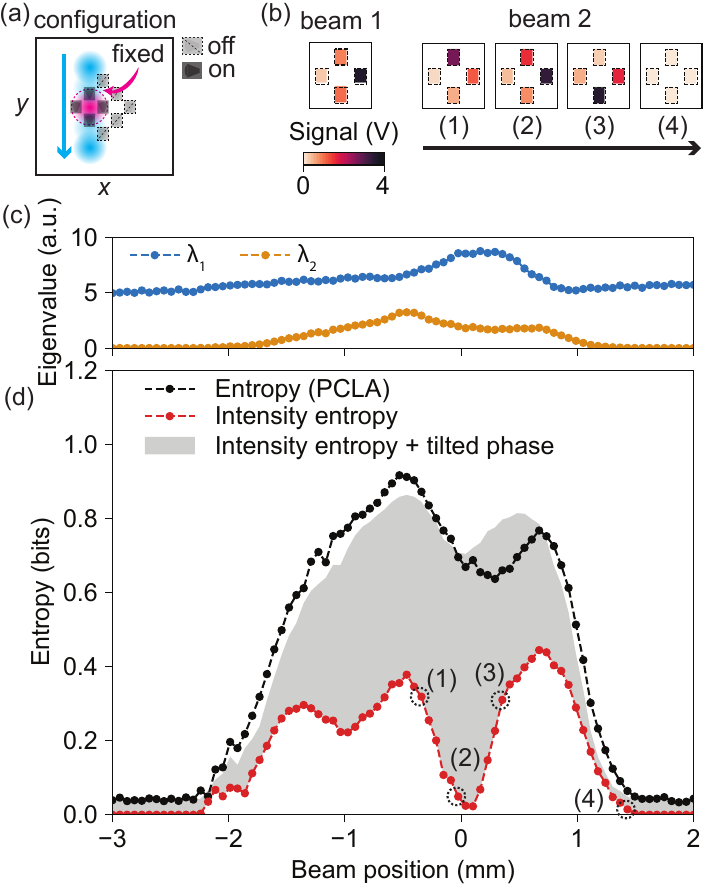}
  \caption{\small \textbf{Measuring two-beam entropy via coherence tomography.} (a) Two-beam configuration: beam 1 is fixed and overlaps with the bottom four grating couplers; beam 2 moves along the $x$ direction. (b) Intensity distribution over a set of four grating couplers for 4 specific positions of beam 2. (c) Measured eigenvalues, output of layers 1 and 2, respectively. (d) Measured entropy. The grey shaded area corresponds to a calculated entropy from the intensity entropy, where a (fitted) phase front tilt of $4.40\deg$ is added to beam 2.}
    \label{fig:3}
    \vspace{-0.3cm}
\end{figure}

\begin{figure*}
\centering
  \includegraphics[scale=0.7]{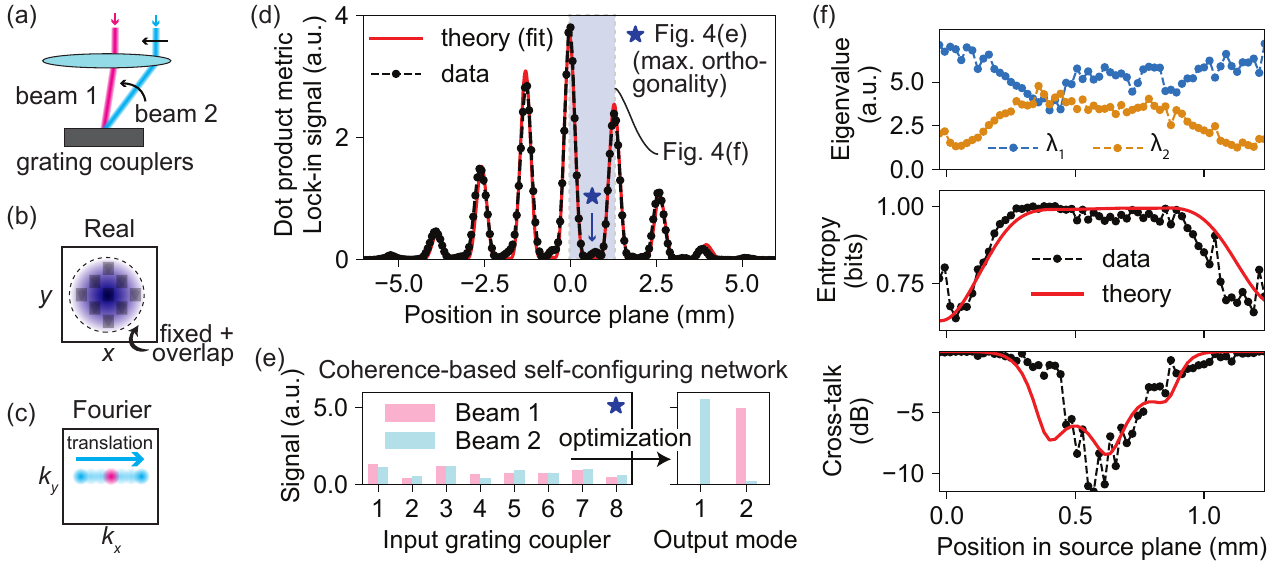}
  \caption{\small \textbf{Discriminating overlapping beams and coherence-based self-configuring networks with PCLA.} (a) Schematic of the beam configuration showing the connection between beam position in the source plane to angle of incidence in this experiment. (b) Beams 1 and 2 are overlapping in real space (on the grating coupler array). (c) Beam 1 is incident at a fixed angle, and beam 2 is swept across Fourier space (varying position in source plane, corresponding to varying the angle of incidence on the chip). (d) Dot product metric between beam 1 and 2, after learning beam 1 with layer 1. (e) PCLA operation at the position of maximum orthogonality (blue star in (d)). The two beams are mutually incoherent and orthogonal, and are separated into two output modes of the network (with cross-talk of -13.0~dB). (f) Fourier-domain sweep. At each point, a 2-layer, sequential optimization is performed. Top row: Measured eigenvalues, output of layers 1 and 2, respectively;  middle row: Entropy of the coherency matrix; bottom row: cross-talk after optimization.}
    \label{fig:4}
    \vspace{-0.3cm}
\end{figure*}

\subsection*{Measuring the entropy of partially coherent light on chip}

We next demonstrate two proof-of-concept applications of the PCLA for coincident beam detection and analysis. \Cref{fig:3} illustrates a set of beam-overlap configurations in which one beam is fixed while a second beam is translated across the grating-coupler array. Depending on the relative position, the two beams exhibit indistinguishable intensity distributions (position~(2)), partially distinguishable intensities (positions~(1) and~(3)), or coupling of only a single beam (position~(4)). Importantly, even when the intensity profiles are indistinguishable, the beams may remain distinguishable in phase profile -- for example, if they impinge on the chip at slightly different angles -- so that beams in position~(2) are not necessarily indistinguishable when phase information is taken into account.
To quantify the information content of the input beams as detected by the PCLA, we introduce the coherency entropy~\cite{hashemi2026chip, gamo1964iii}
\begin{equation}
S(\rho_\text{in}) = -\sum_i \lambda_i \log_2 \lambda_i,
\end{equation}
where $\lambda_i=\pi_i/\sum_j \pi_j$ are the normalized eigenvalues of the input coherency matrix $\rho_\text{in}$. For comparison, we also define an intensity-only entropy $S_I$, which neglects phase correlations and estimates the information accessible from intensity measurements alone (e.g., from the intensity distribution measured at the grating couplers). As shown in Section~S1 of the SI, these experimental observables satisfy $S \geq S_I$ when illuminating the chip with two beams, reflecting the intuitive fact that phase-sensitive measurements can access strictly more information than intensity-only detection. The PCLA enables direct measurement of both quantities.

At position~(2), the intensity-only entropy $S_I \approx 0$ indicates indistinguishable intensity distributions, whereas the full coherency entropy $S \approx 0.6$ reveals the presence of two distinct modes arising from phase differences. The entropy is maximized near position~(1), approaching the value expected for two orthogonal modes with equal weights ($S=1$). To confirm that the additional information captured by $S$ relative to $S_I$ originates from phase structure, we model the intensity-only distribution augmented by a fitted phase tilt between the arrival angles of the two beams of $4.40\deg$, which reproduces the principal features of the measured coherency entropy. These results demonstrate that the PCLA captures information beyond the intensity-only measurements of conventional focal-plane arrays~\cite{brady2025interferometric}.



\subsection*{Coherence-based self-configuring networks}

We now turn to a second illumination configuration in which both beams are spatially overlapped on the $N=9$ input grating couplers, while the angle of incidence of beam~2 is continuously varied, keeping the two beams overlapped in real space (\cref{fig:4}(a–c)). The chip first enables a direct measurement of the spatial orthogonality between the two beams. The first layer of the network is self-configured to maximize coupling of beam~1~\cite{miller_self-configuring_2013}, which is then turned off. While keeping this configuration of the first layer, beam~2 is turned on, and the output power from the first layer is recorded as the angle of arrival of beam~2 is changed (here by changing its position in a "Fourier" source plane -- see SI, Section S3). This allows us to deduce the relative angle of beam~2 at which the two beams are maximally orthogonal (\cref{fig:4}(d) and SI, Section~S2).



At maximal orthogonality, the two beams are mutually incoherent and orthogonal. When the PCLA is re-run under these conditions, it functions as a self-configuring demultiplexer that routes each beam to a distinct output port, with the most powerful beam emerging from the first layer and the second orthogonal beam from the second layer. In this regime, the input modes ${\ket{e_1}, \ket{e_2}}$ coincide with the coherence modes ${\ket{u_1}, \ket{u_2}}$ of the coherency matrix, so separation arises purely from power optimization, without prior knowledge of their spatial profiles. The measured cross-talk between the separated beams is as low as $-13.0~\si{\decibel}$.

More generally, by sweeping the angle of incidence of beam~2 and repeating the tomography, we extract the eigenvalue distribution (\cref{fig:4}(f), top), along with the corresponding coherency entropy and output cross-talk. The entropy is maximized at the point of orthogonality ($\mathcal{S}=1$) and decreases to $\mathcal{S}\approx0.64$ when the beams are aligned in phase and amplitude, where the chip effectively detects a single dominant mode, with the remaining entropy arising from phase differences, residual incoherent background, and imperfect power confinement across the network. Consistently, the PCLA is able to discriminate between the two overlapping beams peaks at maximal orthogonality, as evidenced by the cross-talk data (\cref{fig:4}(f), bottom). All measured observables are quantitatively captured by a theoretical model describing the coupling of the two beams into the discrete grating-coupler array, with model parameters obtained solely from the orthogonality measurement in \cref{fig:4}(d); further details are provided in Section~S4 of the SI. 

\section{Discussion and Conclusion}

The self-configuring network shown in \cref{fig:4} de-multiplexes spatially orthogonal beams solely on the basis of their mutual incoherence, representing a significant extension of prior self-configuring photonic networks that rely on separating spatially orthogonal coherent modes~\cite{milanizadeh2021coherent}~\footnote{Note too, that this separation does not rely on a calibrated measurement of the partially coherent field; though our method does also formally measure that field if we calibrate the network, the separation of the partially coherent field into coherence modes, does not rely on or require any such calibration. }. 
Because the present approach is driven by global, optimization-based tuning of all layer parameters, it is naturally suited to adaptive tracking in optical communication involving sources with modest or low coherence. While our demonstration focused on shaping coherence in the spatial domain, recent network architectures have been demonstrated to extend our work to temporal coherence shaping~\cite{miller2025universal, valdez2025programmable}. In this context, integrated photonic circuits provide a compelling platform for realizing compact, scalable, and dynamically reconfigurable coherence-aware receivers that can operate directly on partially coherent optical fields~\cite{hashemi2026programmable}. Our optimization framework further confers robustness to optical loss, fabrication imperfections, and slow dynamical drifts in the network, which can be continuously compensated \textit{in situ}. 

In practice, the achievable update rate is governed by a trade-off between control loop bandwidth, gradient signal-to-noise ratio, and actuator dynamics inherent to dithering-based control~\cite{zanetto2021dithering}.
State-of-the-art MZI network controllers already demonstrate sub-millisecond stabilization per interferometer~\cite{martinez2024self, sacchi2025integrated}. Such kilohertz-class update rates are sufficient for real-time mitigation of atmospheric turbulence in free-space optical communication systems~\cite{martinez2024self}.
With application-specific, highly parallel control electronics and faster (e.g., electro-optic) phase shifters, update rates well beyond the kilohertz regime appear feasible.


More broadly, the PCLA illustrates how interferometric networks can be used to access information that is fundamentally hidden to intensity-only measurements. By adaptively learning the coherence-mode basis of an unknown optical field, the device converts phase and mutual coherence into directly measurable power observables, enabling a quantitative increase in accessible information as captured by the coherency entropy $S$. This perspective connects naturally to recent work showing that suitably chosen interferometric measurements may overcome apparent information loss in imaging and sensing, including interferometric focal-plane architectures~\cite{brady2025interferometric} and information-theoretic approaches to imaging~\cite{nair2016interferometric,tsang2019resolving, tsang2016quantum, hradil2021exploring, wadood2021experimental, tsang2019resolving, larson2018resurgence, tham2017beating}. Looking forward, scaling such coherence-aware, self-configuring interferometric processors to higher mode counts and faster control bandwidths could enable adaptive, information-optimal front ends for superresolution imaging~\cite{tsang2016quantum}, coronagraphy~\cite{deshler2025experimental}, and dynamically reconfigurable optical sensing, where the relevant measurement basis is not known \emph{a priori} but must be learned directly from the field coherence itself.

Lastly, the same architecture can be operated in reverse to synthesize spatially partially coherent light: by injecting mutually incoherent fields of the same wavelengths and measured powers into the output ports and running the optimized MZI network backward, the chip generates the previously learned coherence state (or, technically, its phase conjugate -- see experimental results in Section~S5 of the SI). Effectively, the approach has formed and recorded the equivalent of a hologram for each of the coherence modes, one such component per self-configuring layer, allowing the later recreation of an equivalent partially coherent field or its phase conjugate, and with the possibility of similarly extending this approach to temporal partial coherence~\cite{miller2025universal}. Taken together, these results establish a practical framework for the on-chip measurement, processing, and synthesis of partially coherent light, paving the way toward programmable coherence control in classical~\cite{roques-carmes_measuring_2024, miller2025universal, valdez2025programmable} and quantum~\cite{roques2025automated, karnieli2025variational} integrated photonic systems.

\section{Authors contributions}
CRC and DABM initially conceived the project. 
CGV, ARK, ZS, VG, and DABM designed the photonic chip.
CGV, ARK, and GC built the control electronics.
ARK, M\v{S}, and DABM developed and implemented the gradient dithering estimation.
PAM, ARK, CGV, and DABM built the experimental setup.
PAM, AK, DABM, and CRC developed the theoretical and numerical models. 
PAM, DABM, and CRC analyzed the experimental data. 
CRC, DABM, OS, and SF supervised the research. PAM, DABM, and CRC wrote the paper with inputs from all authors.

\section{Competing interests}
The authors declare no competing interest.

\section{Data and code availability statement}
The data and codes that support the plots within this paper and other findings of this study are available from the corresponding authors upon reasonable request. Correspondence and requests for materials should be addressed to crc@ista.ac.at.

\section{Acknowledgments}
The authors would like to thank Christina Spägele, Franti\v{s}ek Je\v{r}\'{a}bek, Francesco Zanetto, Alessandro di Tria, Francesco Morichetti, Andrea Melloni, and Marco Sampietro for useful discussions. 

\section{Funding}
M\v{S} acknowledges financial support from the Austrian Federal Ministry of Labour and Economy, the National Foundation for Research, Technology and Development, the Christian Doppler Research Association, and University of Graz. 
SF acknowledges funding by the Air Force Office of Scientific Research (FA9550-21-1-0312). DABM acknowledges funding by the Air Force Office of Scientific Research (FA9550-21-1-0312,
FA9550-23-1-0307). 
CRC is supported by a Stanford Science Fellowship and acknowledges startup funding from the Institute of Science and Technology, Austria (ISTA). 

\bibliographystyle{ieeetr}
\bibliography{bibliography}
\end{document}